\documentclass{ws-procs9x6}

\begin{document}

\title{Y-type Flux-Tube Formation and 
\\ Gluonic Excitations in Baryons: \\ 
From QCD to Quark Model}

\author{Hideo~Suganuma and Hiroko~Ichie}

\address{Faculty of Science, Tokyo Institute of Technology,\\
Ohokayama 2-12-1, Meguro, Tokyo 152-8551, Japan\\
suganuma@th.phys.titech.ac.jp}

\author{Toru~T.~Takahashi}

\address{Yukawa Institute for Theoretical Physics, Kyoto University,\\
Kitashirakawa-Oiwake, Sakyo, Kyoto 606-8502, Japan}

\maketitle

\abstracts{
Using SU(3) lattice QCD, 
we perform the first systematic study for the ground-state three-quark (3Q) potential $V_{\rm 3Q}^{\rm g.s.}$
and the 1st excited-state 3Q potential $V_{\rm 3Q}^{\rm e.s.}$, {\it i.e.}, 
the energies of the ground state and the 1st excited state 
of the gluon field in the presence of the static three quarks.
From the accurate and thorough calculation for 
more than 300 different patterns of 3Q systems, the static ground-state 3Q potential 
$V_{\rm 3Q}^{\rm g.s.}$ is found to be well described 
by the Coulomb plus Y-type linear potential, {\it i.e.}, Y-Ansatz, within 1\%-level deviation.
As a clear evidence for Y-Ansatz, 
Y-type flux-tube formation is actually observed on the lattice in maximally-Abelian projected QCD.
For more than 100 patterns of 3Q systems, 
we calculate the 1st excited-state 3Q potential $V_{\rm 3Q}^{\rm e.s.}$ in quenched lattice QCD, and find 
the gluonic excitation energy $\Delta E_{\rm 3Q} \equiv V_{\rm 3Q}^{\rm e.s.}-V_{\rm 3Q}^{\rm g.s.}$ 
to be about 1 GeV. This large gluonic-excitation energy is conjectured to ensure 
the success of the quark model for the low-lying hadrons 
even without gluonic excitations.
}

\vspace{-0.5cm}

\section{Introduction $\sim$ Hadron Physics based on QCD}

Quantum chromodynamics (QCD), the SU(3) gauge theory,  
was first proposed by Yoichiro Nambu\cite{N66} in 1966 as a candidate for 
the fundamental theory of the strong interaction, 
just after the introduction of the ``new" quantum number, ``color".\cite{HN65} 
In spite of its simple form, 
QCD creates thousands of hadrons and makes the vacuum nontrivial, 
which leads to various interesting nonperturbative phenomena 
such as color confinement\cite{conf2000} and dynamical chiral-symmetry breaking.\cite{NJL61}
This miracle of QCD is due to its strong-coupling nature in the infrared region, 
but the strong-coupling nature itself makes very difficult to deal with QCD.  

In recent years, according to the remarkable progress of the computational power, 
the lattice QCD Monte Carlo calculation becomes a reliable and useful method
for the analysis of nonperturbative QCD,\cite{R97} 
which indicates an important direction in the hadron physics.
In this paper, using lattice QCD, we study the three-quark potential in detail.\cite{TMNS01,TSNM02,TS03,TMNS99}

\section{The Ground-State Three-Quark Potential in QCD}

In general, the three-body force is regarded as a residual interaction in most fields in physics.
In QCD, however, the three-body force among three quarks is 
a ``primary" force reflecting the SU(3) gauge symmetry.
In fact, the three-quark (3Q) potential is directly responsible 
for the structure and properties of baryons, 
similar to the relevant role of the Q-$\bar{\rm Q}$ potential for meson properties, 
and both the Q-$\bar{\rm Q}$ potential and 
the 3Q potential are equally important fundamental quantities in QCD.
Furthermore, the 3Q potential is the key quantity to clarify the quark confinement in baryons.
However, in contrast with a number of studies on the Q-$\bar{\rm Q}$ potential using lattice QCD,\cite{R97,JKM03} 
there was almost no lattice QCD study for the 3Q potential before our study in 1999,\cite{TMNS99}
in spite of its importance in the hadron physics. 

\subsection{Theoretical Form for the 3Q Potential $\sim $ Y-Ansatz}

From the detailed studies with lattice QCD,
the Q-$\bar {\rm Q}$ potential is known to be well described with 
the inter-quark distance $r$ as~\cite{R97,TSNM02} 
\begin{eqnarray}
V_{\rm Q \bar Q}(r)=-\frac{A_{\rm Q\bar Q}}{r}+\sigma_{\rm Q \bar Q}r+C_{\rm Q\bar Q}.
\label{eqn:QQpotential}
\end{eqnarray}
As for the 3Q potential form, there are two theoretical arguments 
for the limits of short and long distances.
\begin{enumerate}
\item
At the short distance, perturbative QCD is applicable,
and therefore 3Q potential is expressed as the sum of the two-body Coulomb potential 
originating from the one-gluon-exchange process.
\item
At the long distance, the strong-coupling expansion of QCD is plausible, and it 
leads to the flux-tube picture.\cite{KS75}
For the 3Q system, there appears a junction which connects the three flux-tubes from the three quarks, 
and Y-type flux-tube formation is expected.\cite{CI86,FRS91,BPV95}
\end{enumerate}

\noindent
Then, we theoretically conjecture the functional form of the 3Q potential 
as the Coulomb plus Y-type linear potential, {\it i.e.,} Y-Ansatz,
\begin{eqnarray}
V_{\rm 3Q}^{\rm g.s.}=-A_{\rm 3Q}\sum_{i<j}\frac1{|r_i-r_j|}+
\sigma_{\rm 3Q}L_{\rm min}+C_{\rm 3Q},
\label{eqn:3Qpotential}
\end{eqnarray}
where $L_{\rm min}$ is the minimal value of the total flux-tube length given by
\begin{eqnarray}
&&L_{\rm min}=\overline{\rm AP}+\overline{\rm BP}+\overline{\rm CP}\\
&=&\Big(\frac{1}{2}(a^2+b^2+c^2)+
\frac{\sqrt{3}}{2}\sqrt{(-a+b+c)(a-b+c)(a+b-c)(a+b+c)}\Big)^{1/2} \nonumber
\end{eqnarray}
as shown in Fig.1.

\begin{figure}[hb]
\vspace{-0.8cm}
\centering
\includegraphics[height=2.8cm]{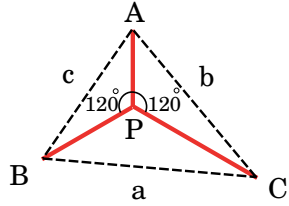}
\vspace{-0.2cm}
\caption{The flux-tube configuration of the 3Q system with the minimal value $L_{\rm min}$ of 
the total flux-tube length. There appears a physical junction linking 
the three flux-tubes at the Fermat point P, and one finds $L_{\rm min}=\overline{\rm AP}+\overline{\rm BP}+\overline{\rm CP}$.
}
\label{fig1}
\vspace{-0.3cm}
\end{figure}

\noindent
Of course, it is nontrivial that these simple arguments on UV and IR limits of QCD hold for the intermediate region as 
$0.2 {\rm fm} < r <1 {\rm fm}$. 
Then, we study the 3Q potential in lattice QCD.
Note that the lattice QCD data itself is completely independent of any Ansatz for the potential form.

\vspace{-0.33cm}

\subsection{The Three-Quark Wilson Loop}

Similar to the Q-$\bar {\rm Q}$ potential calculated with the Wilson loop, 
the 3Q potential can be calculated  with 
the 3Q Wilson loop\cite{TMNS01,TSNM02,TS03,FRS91,BPV95} 
defined on the contour of three large staples as 
\begin{eqnarray}
W_{\rm 3Q}\equiv \frac1{3!}\epsilon_{abc}\epsilon_{a'b'c'}U_1^{aa'}U_2^{bb'}U_3^{cc'}
\end{eqnarray}
with
$U_k\equiv P\exp\{ig\int_{\Gamma_k}dx_\mu A^\mu(x)\}$ in Fig.2.
The 3Q Wilson loop physically means that
a color-singlet gauge-invariant 3Q state is created at $t=0$ and 
is annihilated at $t=T$ with the three quarks spatially fixed for $0<t<T$.

\begin{figure}[hb]
\vspace{-0.27cm}
\centering
\includegraphics[height=3.4cm]{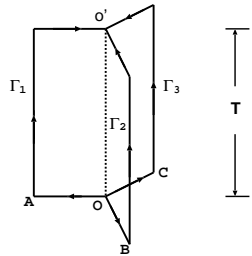}
\vspace{-0.2cm}
\caption{The 3Q Wilson loop.
A color-singlet (gauge-invariant) 3Q state is created at $t=0$ and 
is annihilated at $t=T$. The three quarks are spatially fixed for $0<t<T$.
}
\label{fig2}
\end{figure}

The vacuum expectation value of the 3Q Wilson loop is expressed as 
\begin{eqnarray}
\langle W_{\rm 3Q} \rangle =\sum_{n=0}^{\infty} C_n \exp(-V_n T),
\end{eqnarray}
where $V_n$ denotes the $n$-th energy of the gauge-field configuration 
in the presence of the spatially-fixed three quarks.\cite{TMNS01,TSNM02,TS03}

It is worth mentioning that, while $V_n$ depends only on the 3Q location,
$C_n$ depends on the operator choice at $t=0$ and $T$, {\it e.g.}, path linking between 3Q.
In fact, the gauge-invariant 3Q state prepared at $t=0$ generally includes  
excited-state contributions.
In principle, by taking the large $T$ limit, the ground-state potential $V_0$ can be extracted as 
$
\langle W_{\rm 3Q} \rangle \sim  C_0 \exp(-V_0 T). 
$
However, the large-$T$ limit calculation is difficult in the practical lattice-QCD calculation, 
since the signal decreases exponentially with $T$.
Therefore, for the accurate calculation, it is desired to 
reduce the excited-state component in the 3Q state prepared at $t=0$ and $T$.

\vspace{-0.2cm}

\subsection{Smearing Method: Excited-State Component Reduction}

The smearing method is one of the most popular and useful 
techniques to extract the ground state in lattice QCD,\cite{TSNM02} 
and is actually successful for the ground-state Q-$\bar {\rm Q}$ potential.\cite{TSNM02}
The standard smearing for link-variables is expressed as the 
iterative replacement of the spatial link-variable $U_i (s)$ 
($i=1,2,3$) by the obscured link-variable 
$\bar U_i (s) \in {\rm SU(3)}$ which maximizes 
\begin{equation}
{\rm Re} \,\,{\rm tr} \Big[ \bar U_i^{\dagger}(s) 
\Big\{ \alpha U_i(s)+\sum_{j \ne i, \pm} 
U_{\pm j}(s)U_i(s \pm \hat j)U_{\pm j}^\dagger (s+\hat i) \Big\} \Big]
\end{equation}
with the smearing parameter $\alpha \in {\bf R}$.
Here, we define $U_{-\mu}(s)\equiv U_\mu^\dagger (s-\hat \mu)$.
This procedure is schematically illustrated in Fig.3.
The $n$-th smeared link-variables $U_\mu^{(n)}(s)$ 
are iteratively defined as
$
U_i^{(n)}(s) \equiv \bar U_i^{(n-1)}(s) (i=1,2,3)
$
and 
$
U_4^{(n)}(s) \equiv U_4(s)
$
starting from $U_\mu^{(0)}(s) \equiv U_\mu(s)$. 

\begin{figure}[hb]
\vspace{-0.5cm}
\centering
\includegraphics[height=2cm]{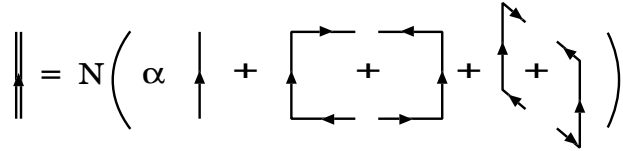}
\vspace{-0.3cm}
\caption{The schematic illustration for the smearing method for the link-variables.}
\vspace{-0.4cm}
\label{fig3}
\end{figure}

Note that the smearing is just a method to 
construct the operator, and hence 
it never changes the physics itself such as the gauge configuration, 
unlike the cooling.
As an important feature, 
the gauge-transformation property of 
$U_\mu^{(n)}(s)$ is just the same as that of  
$U_\mu(s)$, which ensures 
the gauge invariance of the $n$-th smeared 3Q Wilson loop
$\langle W_{\rm 3Q}[U_\mu^{(n)}(s)]\rangle$.

The fat link-variable $U_\mu^{(n)}(s)$ includes a spatial extension 
in terms of the original link-variable $U_\mu(s)$, 
and the smeared ``line'' expressed with $U_\mu^{(n)}(s)$ physically 
corresponds to a Gaussian-distributed ``flux-tube''.\cite{TSNM02}
Therefore, the properly smeared line 
is expected to resemble the ground-state flux-tube. 
Here, the smearing parameter $\alpha$ and 
the iteration number $n$ can be regarded as the 
variational parameters to enhance the ground-state overlap.

Thus, through the selection of the properly smeared 3Q Wilson loop $\langle W_{\rm 3Q}[U_\mu^{(n)}(s)]\rangle$, 
we can construct the ground-state-dominant 3Q operator  
for the accurate measurement of 
the ground-state 3Q potential.\cite{TMNS01,TSNM02,TS03}

\vspace{-0.2cm}

\subsection{Lattice QCD results for the Ground-State 3Q Potential}

For more than 300 different patterns of spatially-fixed 3Q systems, 
we perform the thorough calculation  
of the ground-state potential $V_{\rm 3Q}^{\rm g.s.}$ 
in SU(3) lattice QCD with the standard plaquette action 
with $12^3\times 24$ at $\beta=5.7$ and 
with $16^3\times 32$ at $\beta$=5.8 and 6.0 at the quenched level.
For the accurate measurement, we 
use the smearing method and construct the ground-state-dominant 
3Q operator.\cite{TMNS01,TSNM02,TS03}

To conclude, we find that the static ground-state 3Q potential $V_{\rm 3Q}^{\rm g.s.}$
is well described by the Coulomb plus Y-type linear potential, {\it i.e.}, Y-Ansatz, 
within 1\%-level deviation,\cite{TMNS01,TSNM02} as shown in Table 1.

\begin{table}[htb]
\vspace{-0.25cm}
\tbl{
Examples of the 3Q potential $V_{\rm 3Q}^{\rm latt}$ for the 3Q system 
put on $(i,0,0)$, $(0,j,0)$, $(0,0,k)$ in ${\bf R}^3$ 
in lattice QCD at $\beta$=6.0.
For each 3Q configuration, 
$V_{\rm 3Q}^{\rm latt}$ is measured 
from the single- exponential fit as 
$\langle W_{\rm 3Q}\rangle=\bar{C}e^{-V_{\rm 3Q}T}$.
$\bar C \simeq 1$ physically means the ground-state dominance in the smeared 3Q Wilson loop. 
We add the difference from the best-fit Y-Ansatz, 
$\Delta V \equiv |V^{\rm latt}_{\rm 3Q}-V^{\rm Y}_{\rm 3Q}|$, 
which is only about 1\% of the typical scale of $V_{\rm 3Q}$.
The listed value is measured in the lattice unit.
\vspace{-0.2cm}
\vspace*{1pt}}
{\footnotesize
\newcommand{\m}{\hphantom{$-$}}
\begin{tabular}{c c c c c c c c} \hline\hline
$(i, j, k)$ & $V_{\rm 3Q}^{\rm latt}$
& \lower.4ex\hbox{$\bar{C}$} & 
$\Delta V$ &
$(i, j, k)$ & $V_{\rm 3Q}^{\rm latt}$
& \lower.4ex\hbox{$\bar{C}$} & 
$\Delta V$ \\
\hline
 (0,1,1) &  0.6778(~6)  &  0.9784( 24)  & 0.0012      &
 (0,2,3) &  1.0259(24)  &  0.9607( 91)  & 0.0045      \\
 (0,1,2) &  0.8234(11)  &  0.9712( 45)  & 0.0042      &
 (0,2,4) &  1.0946(32)  &  0.9657(120)  & 0.0003      \\
 (0,1,3) &  0.9183(17)  &  0.9769( 65)  & 0.0045      &
 (0,2,5) &  1.1454(41)  &  0.9282(149)  & 0.0064      \\
 (0,1,4) &  0.9859(24)  &  0.9589( 92)  & 0.0050      &
 (0,2,6) &  1.2075(28)  &  0.9464( 76)  & 0.0018      \\
 (0,1,5) &  1.0463(30)  &  0.9495(112)  & 0.0064      &
 (0,2,7) &  1.2563(33)  &  0.9262( 90)  & 0.0012      \\
 (0,1,6) &  1.1069(40)  &  0.9595(152)  & 0.0122      &
 (0,3,3) &  1.0999(23)  &  0.9566( 62)  & 0.0031      \\
 (0,1,7) &  1.1572(50)  &  0.9374(192)  & 0.0102      &
 (0,3,4) &  1.1595(25)  &  0.9454( 67)  & 0.0044      \\
 (0,2,2) &  0.9430(21)  &  0.9586( 78)  & 0.0095      &
 (0,3,5) &  1.2170(25)  &  0.9426( 65)  & 0.0026      \\
\hline\hline
\end{tabular}
\label{table1} }
\end{table}

\vspace{-0.35cm}

We summarize in Table 2 the lattice QCD results 
for the string tension and the Coulomb coefficient, 
with comparing between 3Q and Q-$\bar{\rm Q}$ potentials.
As remarkable features, 
we find the universality of the string tension between 
the 3Q and Q-$\bar {\rm Q}$ systems and 
the one-gluon-exchange result as~\cite{TMNS01,TSNM02}
\begin{eqnarray}
\sigma_{\rm 3Q}\simeq\sigma_{\rm Q\bar Q}, \qquad A_{\rm 3Q}\simeq\frac12 A_{\rm Q\bar Q}.
\end{eqnarray}

\begin{table}[htb]
\vspace{-0.5cm}
\newcommand{\cc}[1]{\multicolumn{1}{c}{#1}}
\tbl{
The best-fit parameter set $(\sigma, A, C)$ in the function form of Y-Ansatz,  
$V_{\rm 3Q}=-A_{\rm 3Q}\sum_{i<j}\frac{1}{|{\bf r}_i-{\bf r}_j|}
+\sigma_{\rm 3Q}L_{\rm min}+C_{\rm 3Q}$, 
where $L_{\rm min}$ denotes the minimal value of the Y-type flux-tube length.
The similar fit on  
the Q-$\bar{\rm Q}$ potential (on-axis) is also listed.
The physical unit is determined so as to reproduce $\sqrt{\sigma}$=427MeV 
for the Q-$\bar{\rm Q}$ potential.
The universality of the string tension
and the OGE result on the Coulomb coefficient are found as 
$\sigma_{\rm 3Q} \simeq \sigma_{\rm Q\bar{\rm Q}}$ and
$A_{\rm 3Q} \simeq \frac12 A_{\rm Q\bar{\rm Q}}$, respectively.
\vspace*{1pt}}
{\footnotesize
\begin{tabular}{llllll} \hline \hline
 &  & \cc{$\sigma$~[$a^{-2}$]~~} & \cc{$\sqrt{\sigma}$~[MeV]} & \cc{$A$~~~~} & \cc{$C$~[$a^{-1}$]~~} 
\\ \hline
$\beta =5.7$ & ${\rm 3Q_{Y}}$ & 
              $0.1524(28)$ & \cc{$413.0$} & $0.1331(66)$ & $0.9182(213)$ 
\\ 
($a \simeq 0.186{\rm fm}$) & ${\rm Q\bar{Q}}$ & 
              $0.1629(47)$ & \cc{$427~~$} & $0.2793(116)  $ & $0.6203(161)$ 
\\ \hline
$\beta =5.8$ & ${\rm 3Q_{Y}}$ & 
              $0.1027(6)$ & \cc{$416.6$} & $0.1230(20)$ & $0.9085(55)$ 
\\
($a \simeq 0.152{\rm fm}$) & ${\rm Q\bar{Q}}$ & 
              $0.1079(28)$ & \cc{$427~~$} & $0.2607(174)  $ & $0.6115(197)$ 
\\ \hline
$\beta =6.0$ & ${\rm 3Q_{Y}}$ & 
              $0.0460(4)$ & \cc{$407.1$} & $0.1366(11)$ & $0.9599(35)$ 
\\ 
($a \simeq 0.104{\rm fm}$) & ${\rm Q\bar{Q}}$ & 
              $0.0506(7)$ & \cc{$427~~$} & $0.2768(24)  $ & $0.6374(30)$
\\ 
\hline\hline
\end{tabular}
\label{Table2} }
\end{table}

\vspace{-0.55cm}

\subsection{Further Investigations with various Ans\"atze}

\noindent
{\bf Lattice Coulomb plus Y-type Linear Ansatz}~: 
Considering the lattice discretization error at a very short distance,
we perform also more accurate fitting 
with the lattice Coulomb plus Y-type linear potential.
We obtain almost the same result and confirm that Y-Ansatz is correct.\cite{TSNM02}

\vspace{0.2cm}

\noindent
{\bf Generalized Y-type Linear Ansatz}~: 
Considering a possible contamination of the $\Delta$-type flux at a short distance, 
we study generalized Y-Ansatz,\cite{TSNM02} which includes both 
Y and $\Delta$ Ans\"atze in some limits.
The flux-tube core radius $R$ is found to be rather small as $R\simeq$ 0.08fm, which 
again supports Y-Ansatz.\cite{TSNM02} (See Fig.4.)

\begin{figure}[hb]
\vspace{-0.6cm}
\centering
\includegraphics[height=3cm]{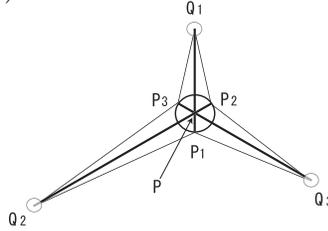}
\vspace{-0.15cm}
\caption{ 
Generalized Y-Ansatz.
The three quarks are spatially fixed at ${\rm Q}_1$, ${\rm Q}_2$, ${\rm Q}_3$.
The point ${\rm P}_1$ is taken inside the circle 
around the physical junction (Fermat point) P 
so as to minimize 
${\rm P}_1{\rm Q}_2+{\rm P}_1{\rm Q}_3$. ${\rm P}_2$ and ${\rm P}_3$ 
are similarly defined.
In generalized Y-Ansatz, 
$
\overline{L_{\rm min}} \equiv \frac12 
(\overline{\rm P_1Q_2} + \overline{\rm P_1Q_3} + \overline{\rm P_2Q_1} +
 \overline{\rm P_2Q_3} + \overline{\rm P_3Q_1} + \overline{\rm P_3Q_2})
$
is used instead of $L_{\rm min}$
}
\label{fig4}
\end{figure}

\vspace{-0.35cm}

\noindent
{\bf Yukawa plus Y-type Linear Ansatz}~: 
As other type of generalization, 
we investigate also the Yukawa plus Y-type linear potential,
\begin{eqnarray}
V_{\rm 3Q}=-A_{\rm 3Q}^{\rm Yukawa} \sum_{i<j}\frac{e^{-m|r_i-r_j|}}{|r_i-r_j|}+
\sigma_{\rm 3Q}^{\rm Yukawa}L_{\rm min}+C_{\rm 3Q}^{\rm Yukawa},
\end{eqnarray}
as is indicated in the dual superconductor picture for color confinement.\cite{conf2000,SST95}
However, we obtain $m \simeq 0$ for the screening mass $m$, 
which would support the type-II dual superconductor,\cite{SST95} if this picture is correct.
In any case, the Coulomb plus Y-type linear potential 
is confirmed once again.\cite{TSNM02}

\vspace{-0.4cm}

\subsection{Other Recent Studies on the 3Q Potential}

To clarify the current status of the 3Q potential, 
we introduce two recent studies on the 3Q potential.

\vspace{0.2cm}

\noindent
{\bf de Forcrand's group}~: 
Recently, de Forcrand's group, who supported $\Delta$-Ansatz in lattice QCD,\cite{AdFT02} 
seems to change their opinion from $\Delta$-Ansatz to Y-Ansatz\cite{JdF03} 
except for a very short distance, 
where the linear potential seems negligible compared with the Coulomb contribution.
(As a problem of their argument, 
they relied on the continuum Coulomb potential 
even for the subtle argument at the very short distance, 
where the lattice Coulomb potential should be used.)

\vspace{0.2cm}

\noindent
{\bf Cornwall}~: 
One of the theoretical basis of $\Delta$-Ansatz was Cornwall's conjecture 
based on the vortex vacuum model.\cite{C96} 
Very recently, motivated by our studies, 
Cornwall re-examined his previous work and found an 
error in his model calculation. 
His corrected answer is Y-Ansatz instead of $\Delta$-Ansatz.\cite{C03}

\vspace{0.2cm}

In this way, Y-Ansatz for the static 3Q potential
seems almost settled both in lattice QCD and in analytic framework. 

\vspace{-0.4cm}

\section{Y-type Flux-Tube Formation in Lattice QCD}

Recently, as a clear evidence for Y-Ansatz, 
Y-type flux-tube formation is actually observed 
in the maximally-Abelian projected QCD 
from the direct measurement for the action density of the gluon field 
in the spatially-fixed 3Q system.\cite{IBSS03,TSIMN03,STI03} (See Fig.5)

\begin{figure}[hb]
\vspace{-0.4cm}
\centering
\includegraphics[width=4.5in]{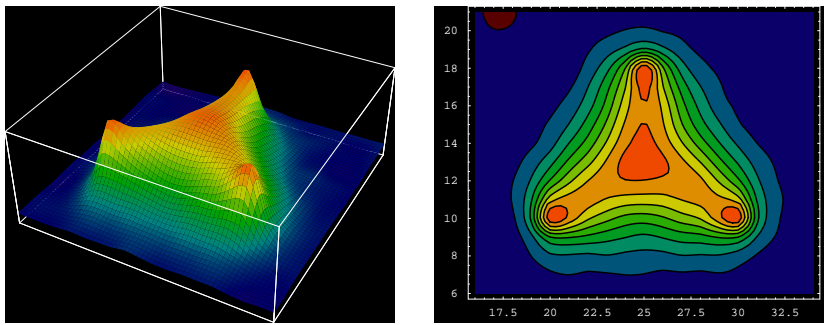}
\vspace{-0.6cm}
\caption{
The lattice QCD result for Y-type flux-tube formation 
in the spatially-fixed 3Q system 
in maximally-Abelian projected QCD.
The distance between the junction and each quark is about 0.5 fm.
}
\label{fig5}
\end{figure}

\section{The Excited-State Three-Quark Potential in QCD}

In 1969, Y.~Nambu first pointed out the string picture for hadrons\cite{N69,N70} 
to explain  the Veneziano amplitude\cite{V68} on hadron reactions and resonances.
Since then, the string picture has been one of the most important pictures for hadrons 
and has provided many interesting ideas in the wide region of the elementary particle physics.

For instance, the hadronic string creates infinite number of hadron resonances as the vibrational modes, 
and these excitations lead to the Hagedorn ``ultimate" temperature,\cite{H65} 
which gives an interesting  theoretical picture for 
the QCD phase transition.

For the real hadrons, of course, the hadronic string is to have a spatial extension like the flux-tube, 
as the result of one-dimensional squeezing of the color-electric flux 
in accordance with color confinement.\cite{N74} 
Therefore, the vibrational modes of the hadronic flux-tube should be much more complicated, and 
the analysis of the excitation modes is important to clarify the underlying picture for real hadrons. 

In the language of QCD, such non-quark-origin excitation is called as the ``gluonic excitation", and 
is physically interpreted as the excitation of 
the gluon-field configuration in the presence of the quark-antiquark pair or the three quarks 
in a color-singlet state.

In the hadron physics, the gluonic excitation is one of the interesting phenomena 
beyond the quark model, and relates to the hybrid hadrons such as $q\bar qG$ and $qqqG$. 
In particular, the hybrid meson includes the exotic hadrons with 
$J^{PC}=0^{--},0^{+-},1^{-+},2^{+-},\cdots$,
which cannot be constructed within the simple quark model.

In this section, 
we study the excited-state 3Q potential and the gluonic excitation 
using lattice QCD,\cite{TS03} 
to get deeper insight on these excitations beyond the hypothetical models
such as the string and the flux-tube models.
Here, the excited-state 3Q potential is 
the energy of the excited state of the gluon-field configuration 
in the presence of the static three quarks, and 
the gluonic-excitation energy is expressed as  
the energy difference between the ground-state 3Q potential  
and the excited-state 3Q potential.

\vspace{-0.3cm}

\subsection{Formalism to extract Excited-State 3Q Potentials}

We present the formalism to extract the 
excited-state potential.\cite{TS03} 
For the simple notation, the ground state is regarded as the ``0-th 
excited state''.
For the physical eigenstates of the QCD Hamiltonian $\hat H$ 
for the spatially-fixed 3Q system, we denote 
the $n$-th excited state by $|n \rangle$ ($n=0,1,2,\cdots$).
Since the three quarks are spatially fixed in this case, 
the eigenvalue of $\hat H$ is expressed by a static potential
as $\hat H|n\rangle=V_n|n\rangle$, 
where $V_n$ denotes the $n$-th excited-state 3Q potential. 
Note that both $V_n$ and $|n \rangle $ are universal 
physical quantities relating to the QCD Hamiltonian $\hat H$.
In fact, $V_n$ depends only on the 3Q location, and 
$|n \rangle $ satisfies the orthogonal condition as $\langle m|n \rangle=\delta_{mn}$.

Suppose that $|\Phi_k \rangle \ (k=0,1,2,\cdots)$ are arbitrary given 
independent spatially-fixed 3Q states. 
In general, each 3Q state $|\Phi_k \rangle$ can be expressed with 
a linear combination of the 3Q physical eigenstates 
$|n\rangle $ as 
\begin{eqnarray}
|\Phi_k \rangle =c_0^k|0\rangle+c_1^k|1\rangle+c_2^k|2\rangle+\cdots.
\end{eqnarray}
Here, the coefficients $c_n^k$ depend on the selection of 
$|\Phi_k \rangle$, and hence they are not universal quantities.

The Euclidean-time evolution of the 3Q state $|\Phi_k(t)\rangle$ is 
expressed with the operator $e^{-\hat Ht}$, which corresponds to 
the transfer matrix in lattice QCD. 
The overlap $\langle \Phi_j(T)|\Phi_k(0)\rangle$ is given by 
the 3Q Wilson loop with the initial state $|\Phi_k\rangle$ 
at $t=0$ and the final state $|\Phi_j\rangle$ at $t=T$, 
and is expressed in the Euclidean Heisenberg picture as 
\begin{eqnarray}
W^{jk}_T&\equiv& 
\langle \Phi_j|W_{\rm 3Q}(T)|\Phi_k\rangle
=\langle\Phi_j(T)|\Phi_k(0)\rangle 
=\langle\Phi_j|e^{-\hat HT}|\Phi_k\rangle \nonumber\\
&=&\sum_{m=0}^\infty \sum_{n=0}^\infty \bar c_m^j c_n^k
\langle m|e^{-\hat HT}|n \rangle
=\sum_{n=0}^\infty \bar c_n^j c_n^k e^{-V_nT}.
\end{eqnarray}
Using the matrix $C$ satisfying $C^{nk} =c_n^k$
and the diagonal matrix $\Lambda_T$ as $\Lambda_T^{mn}=e^{-V_nT}\delta^{mn}$, 
we rewrite the above relation as 
\begin{equation}
W_T=C^\dagger \Lambda_T C.
\label{WandLambda}
\end{equation}
Note here that $C$ is not a unitary matrix, and hence this relation 
does not mean the simple diagonalization by the unitary transformation.

Since we are interested in the 3Q potential $V_{n}$ 
in $\Lambda_T$ rather than the non-universal matrix $C$, 
we single out $V_{n}$ from the 3Q Wilson loop $W_T$ as
\begin{eqnarray}
W^{-1}_TW_{T+1}&=&\{C^\dagger \Lambda_T C\}^{-1} C^\dagger \Lambda_{T+1} C
=C^{-1}{\rm diag}(e^{-V_0},e^{-V_1},\cdots)C,~~~
\end{eqnarray}
which is a similarity transformation.
Then, $e^{-V_n}$ can be obtained as the eigenvalues of the matrix 
$W_T^{-1}W_{T+1}$, {\it i.e.}, solutions of the secular equation, 
\begin{eqnarray}
{\rm det}\{W_T^{-1}W_{T+1}-t{\bf 1}\}=\prod_{n}(e^{-V_n}-t)=0.
\label{secular}
\end{eqnarray}
Thus, the 3Q potential $V_n$ 
can be obtained from the matrix  $W_T^{-1}W_{T+1}$.

In the practical calculation, we prepare $N$ independent sample states 
$|\Phi_k \rangle \ (k=0,1,\cdots,N-1)$. 
By choosing appropriate states $|\Phi_k \rangle$ 
so as not to include highly excited-state components, 
the physical states $|n \rangle$ can be truncated as $0\le n \le N-1$.
Then, $W_T$, $C$ and $\Lambda_T$ are reduced into $N\times N$ matrices, 
and the secular equation (\ref{secular}) becomes the $N$-th 
order equation. 

\vspace{-0.5cm}

\subsection{Lattice QCD results for the Excited-State 3Q Potential}

For more than 100 different patterns of spatially-fixed 3Q systems, 
we study the excited-state potential $V_{\rm 3Q}^{\rm e.s.}$ 
using lattice QCD with $16^3\times 32$ at $\beta$=5.8 and 6.0 at the quenched level.\cite{TS03} 
In Fig.6, we show the first lattice QCD results for the 
excited-state 3Q potential $V_{\rm 3Q}^{\rm e.s.}$ as well as 
the ground-state potential $V_{\rm 3Q}^{\rm g.s.}$.
(In Fig.6, the minimal length $L_{\rm min}$ of the Y-type flux-tube  
is used as a label to distinguish the three-quark configuration.)

The energy gap between $V_{\rm 3Q}^{\rm g.s.}$ and $V_{\rm 3Q}^{\rm e.s.}$ physically means 
the excitation energy of the gluon-field configuration in the presence of the spatially-fixed three quarks, 
and the gluonic excitation energy $\Delta E_{\rm 3Q} \equiv V_{\rm 3Q}^{\rm e.s.}-V_{\rm 3Q}^{\rm g.s.}$ 
is found to be about 1GeV or more~\cite{TS03,TSIMN03,STI03}
in the typical hadronic scale as 
$L_{\rm min}\sim 1~{\rm fm}$.

Note that the gluonic excitation energy of about 1GeV is rather large in comparison with 
the excitation energies of the quark origin, and 
such a gluonic excitation would contribute significantly in
the highly-excited baryons with the excitation energy above 1 GeV. 
The present result predicts that the lowest hybrid baryon, which is described as $qqqG$ in
the valence picture, has a large mass of about 2 GeV.\cite{TS03}

Together with the recent lattice result\cite{JKM03} indicating 
the gluonic excitation energy $\Delta E_{\rm Q \bar Q} \equiv V_{\rm Q \bar Q}^{\rm e.s.}-V_{\rm Q \bar Q}^{\rm g.s.}$ 
for the Q-${\rm \bar Q}$ system to be in the order of 1GeV,  
the present result seems to suggest the constituent gluon mass of about 1GeV
in terms of the constituent gluon picture.

\begin{figure}[hb]
\vspace{-0.55cm}
\centerline{
\includegraphics[height=4.8cm]{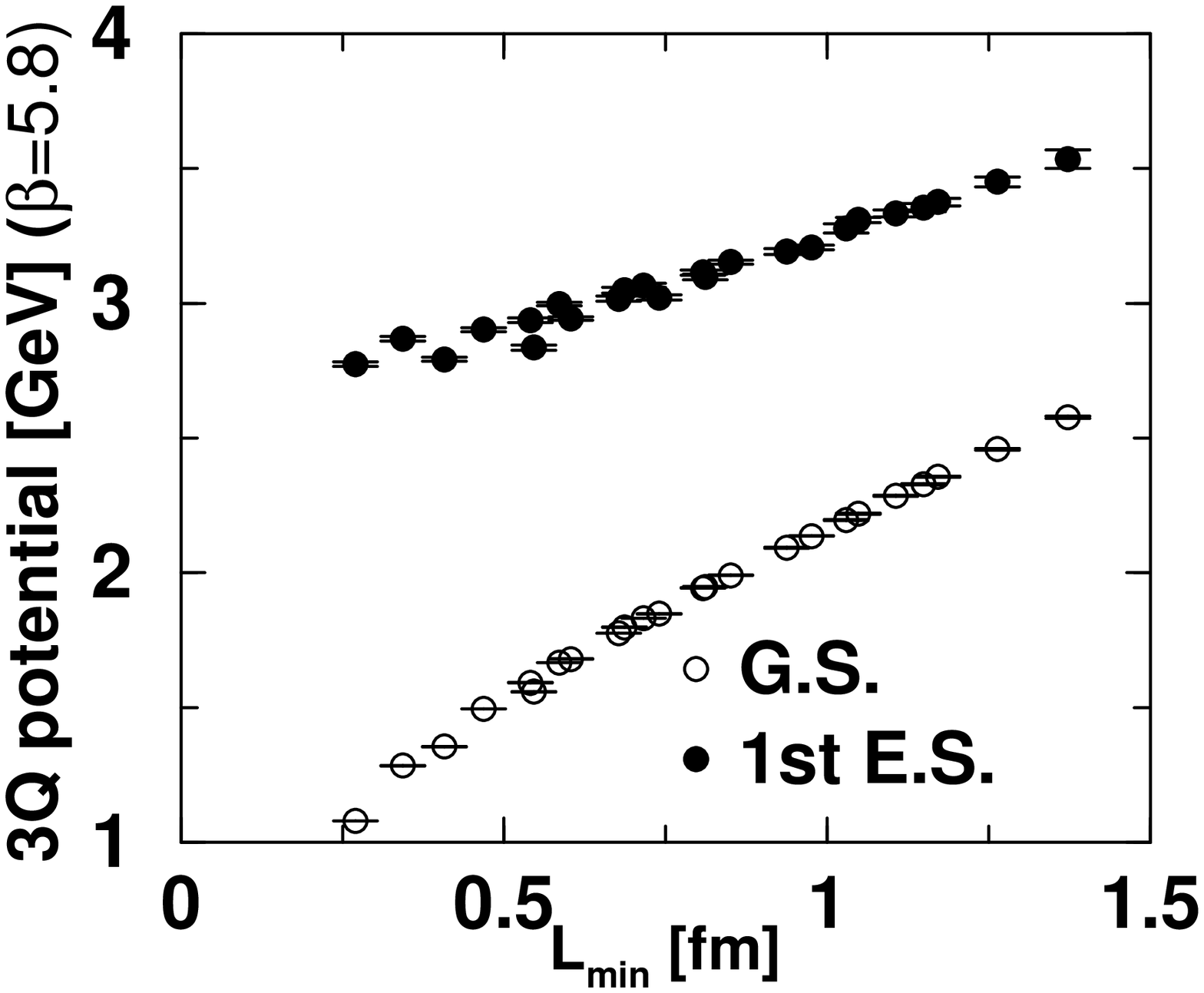}
\includegraphics[height=4.8cm]{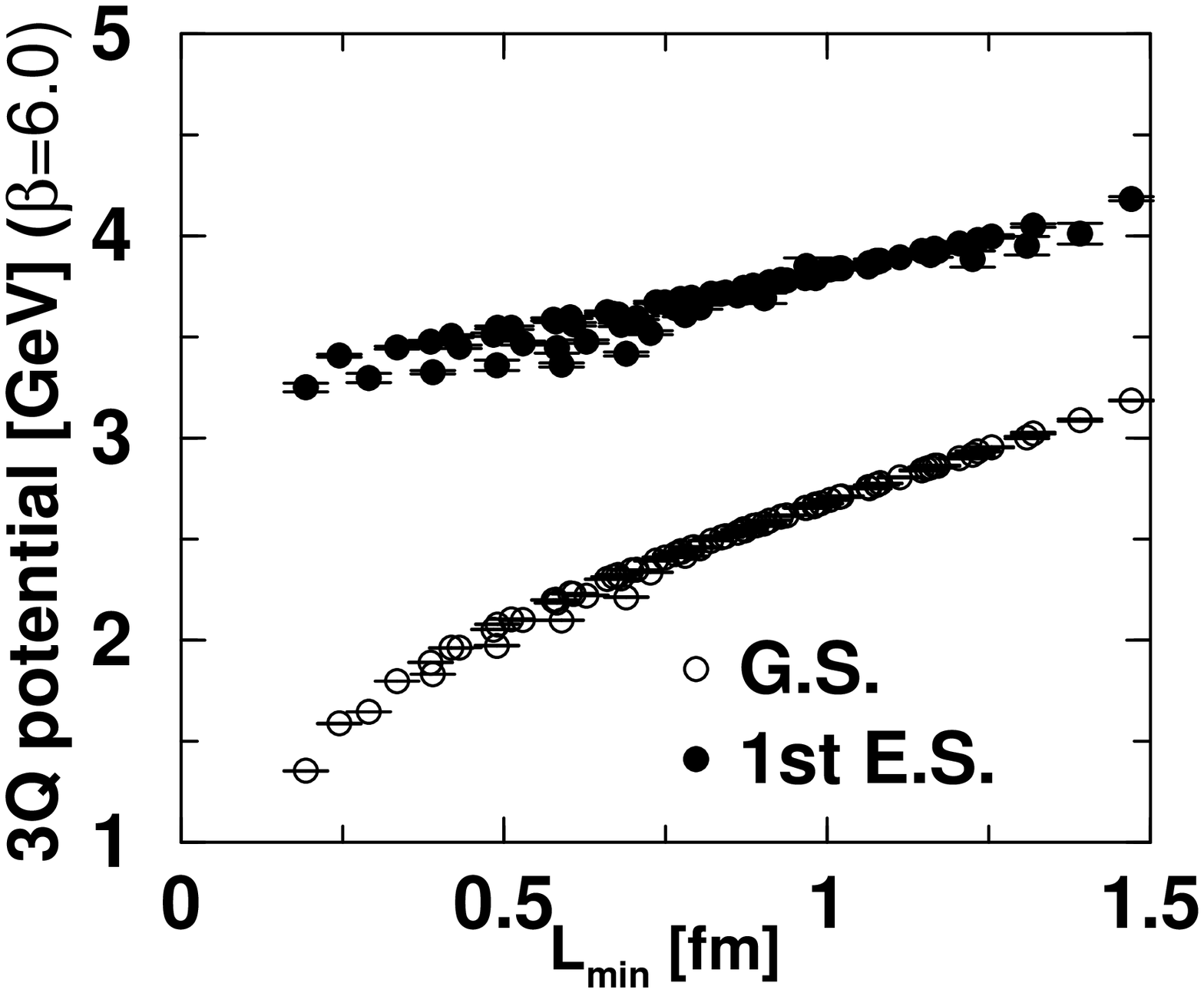}
}
\vspace{-0.35cm}
\caption{
The lattice QCD results of the ground-state 3Q potential 
$V_{\rm 3Q}^{\rm g.s.}$ (open circles) and the 1st excited-state 3Q
potential $V_{\rm 3Q}^{\rm e.s.}$ (filled circles) as the function
of $L_{\rm min}$.
These lattice results at $\beta=5.8$ and $\beta=6.0$ well coincide 
besides an irrelevant overall constant. 
The gluonic excitation energy 
$\Delta E_{\rm 3Q} \equiv V_{\rm 3Q}^{\rm e.s.}-V_{\rm 3Q}^{\rm g.s.}$ 
is found to be about 1GeV in the hadronic scale as $L_{\rm min} \sim 1 {\rm fm}$.
}
\label{fig6}
\end{figure}

\subsection{Functional Form of the Excited-State 
3Q Potential and Comparison with the Q-$\bar Q$ System}

From the lattice QCD data, we attempt to seek the functional form of the excited-state 3Q potential $V_{\rm 3Q}^{\rm e.s.}$, 
but find no simple plausible form of $V_{\rm 3Q}^{\rm e.s.}$, 
unlike the ground-state 3Q potential $V_{\rm 3Q}^{\rm g.s.}$.

Next, we compare the 3Q gluonic excitation with the Q-$\bar {\rm Q}$ gluonic excitation, 
considering the nature of the Y-type junction. 
If the Y-type junction behaves as a quasi-fixed edge of the three flux-tubes, 
these three flux-tubes would behave as independent three Q-$\bar {\rm Q}$ systems, 
and therefore the 3Q gluonic excitation would be approximated as a simple incoherent sum of 
the three Q-$\bar {\rm Q}$ gluonic excitations.
If the Y-type junction behaves as a quasi-free edge, 
the 3Q gluonic-excitation energy $\Delta E_{\rm 3Q}$ would be smaller than 
each of Q-$\bar {\rm Q}$ gluonic-excitation energies $\Delta E_{\rm Q \bar Q}$ 
corresponding to the three flux-tubes, 
since the string with fixed edges has a larger  vibrational energy.

Through the comparison of $V_{\rm 3Q}^{\rm e.s.}$ or $\Delta E_{\rm 3Q}$ with 
several possible linear combinations of $V_{\rm Q\bar Q}^{\rm e.s.}$ or $\Delta E_{\rm Q \bar Q}$,
we find no simple relation between them.
This fact is conjectured to reflect the complicated vibrational mode 
on the Y-type flux-tube, due to the interference among the vibrational modes on the three flux-tubes 
through the junction, which may indicate the quasi-free behavior of the Y-type junction. 

\vspace{-0.3cm}

\section{Behind the Success of the Quark Model}

Finally, we consider the connection between QCD and the quark model 
in terms of the gluonic excitation.\cite{TS03,TSIMN03,STI03} 
While QCD is described with quarks and gluons, 
the simple quark model successfully describes low-lying hadrons 
even without explicit gluonic modes.
In fact, the gluonic excitation seems invisible in the low-lying hadron spectra, 
which is rather mysterious.

On this point, we find the gluonic-excitation energy to be about 1GeV or more, 
which is rather large compared with the excitation energies of the quark origin,
and therefore the effect of gluonic excitations is negligible 
and quark degrees of freedom plays the dominant role 
in low-lying hadrons with the excitation energy below 1GeV.

Thus, the large gluonic-excitation energy of about 1GeV gives the physical reason for 
the invisible gluonic excitation in low-lying hadrons, 
which would play the key role for the success of the quark model 
without gluonic-excitation modes.\cite{TS03,TSIMN03,STI03} 

In Fig.7, by way of the flux-tube picture, 
we present a possible scenario from QCD to the massive quark model 
in terms of color confinement and dynamical chiral-symmetry breaking.\cite{TSIMN03,STI03}

\begin{figure}[hb]
\vspace{-2.8cm}
\centerline{\includegraphics[width=5in]{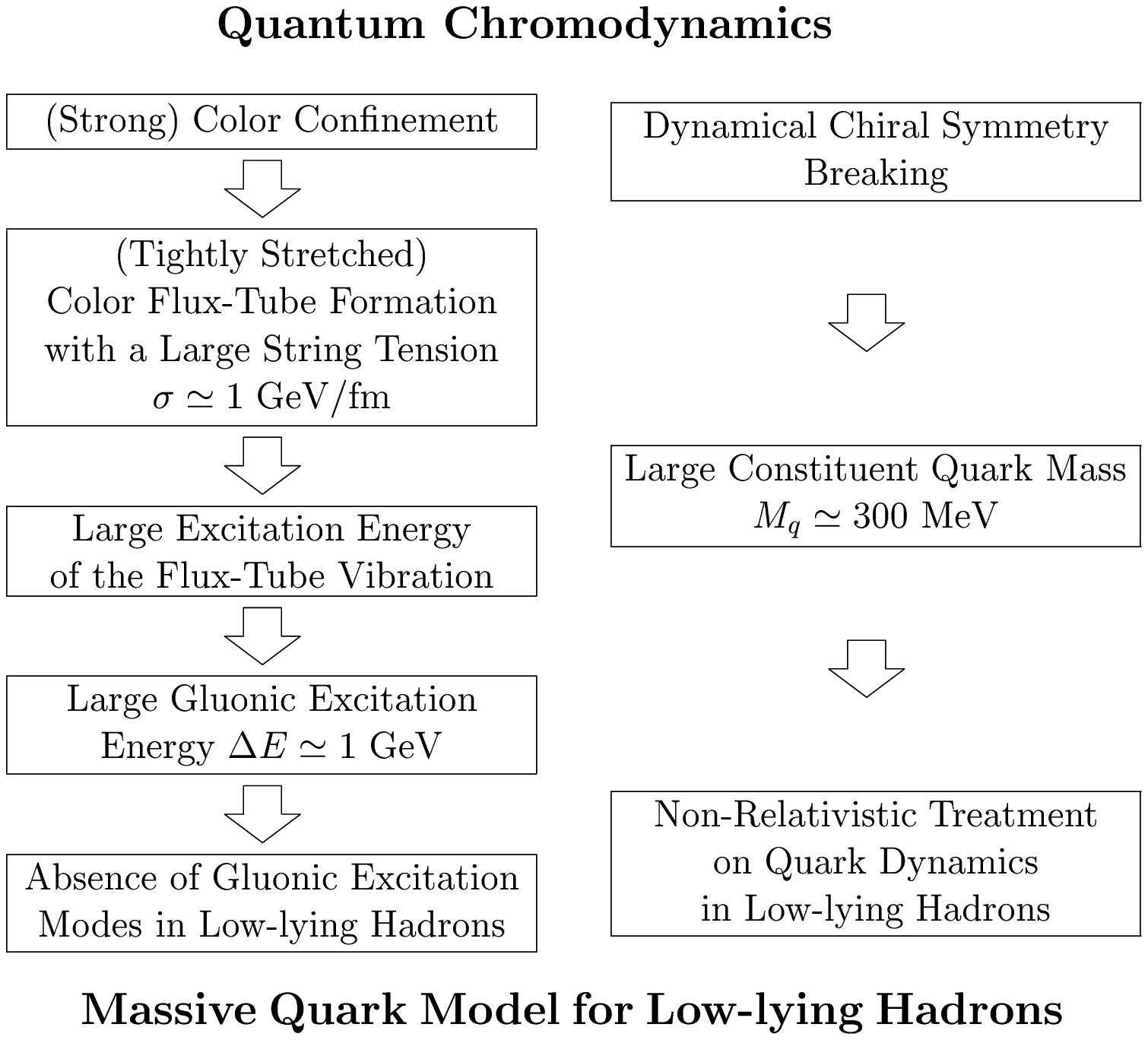}}
\vspace{-7cm}
\caption{A possible scenario from QCD to the quark model in terms of 
color confinement and dynamical chiral-symmetry breaking (DCSB).
DCSB provides a large constituent quark mass of about 300MeV, 
which enables the non-relativistic treatment for quark dynamics. 
Color confinement provides the color flux-tube formation among quarks 
with a large string tension of $\sigma \simeq$ 1 GeV/fm.
In the flux-tube picture, the gluonic excitation is described as the flux-tube vibration, 
and the flux-tube vibrational energy is expected to be large, 
reflecting the large string tension.
The large gluonic-excitation energy of about 1GeV leads to 
the absence of the gluonic mode in low-lying hadrons, 
which would play the key role to the success of the quark model without gluonic excitation modes.
}
\label{fig7}
\end{figure}

\vspace{-0.5cm}

\section{Summary and Concluding Remarks}

Using SU(3) lattice QCD, we have studied the ground-state 3Q potential $V_{\rm 3Q}^{\rm g.s.}$ and 
the 1st excited-state 3Q potential $V_{\rm 3Q}^{\rm e.s.}$. 

From the accurate and thorough calculation for 
more than 300 different patterns of 3Q systems, 
we have found that the static ground-state 3Q potential 
$V_{\rm 3Q}^{\rm g.s.}$ is well described 
by the Coulomb plus Y-type linear potential, {\it i.e.}, Y-Ansatz, within 1\%-level deviation.
As a clear evidence for Y-Ansatz, 
Y-type flux-tube formation has been actually observed 
on the lattice in maximally-Abelian projected QCD.

For more than 100 different patterns of 3Q systems, 
we have performed the first study of 
the 1st excited-state 3Q potential $V_{\rm 3Q}^{\rm e.s.}$ in quenched lattice QCD, 
and have found the gluonic excitation energy 
$\Delta E_{\rm 3Q} \equiv V_{\rm 3Q}^{\rm e.s.}-V_{\rm 3Q}^{\rm g.s.}$ to be about 1 GeV.
This indicates that the hybrid baryons ($qqqG$) are to be rather heavy and 
appear in the spectrum above 2GeV.
We have conjectured that 
the large gluonic-excitation energy of about 1GeV leads to  
the success of the quark model for the low-lying hadrons 
even without gluonic excitations.

\vspace{0.25cm}

\noindent
{\bf Acknowledgements}~~H.S. thanks all the participants of Confinement2003.

\end{document}